# Ultra-fast Real-time Target Recognition Using a Shift, Scale, and Rotation Invariant Hybrid Opto-electronic Joint Transform Correlator

**Xi Shen**[*]
*Department of ECE, Northwestern University, Evanston, IL, 60208, USA*
**Julian Gamboa**[*]
*Department of ECE, Northwestern University, Evanston, IL, 60208, USA*
**Tabassom Hamidfar**
*Department of ECE, Northwestern University, Evanston, IL, 60208, USA*
**Shamima A. Mitu**
*Department of ECE, Northwestern University, Evanston, IL, 60208, USA*
**Selim M. Shahriar**
*Department of ECE, Northwestern University, Evanston, IL, 60208, USA*
*Department of Physics and Astronomy, Northwestern University, Evanston, IL, 60208, USA*
[*]Authors contributed equally to this work

**Abstract:** Hybrid Opto-electronic correlators (HOC) overcome many limitations of all-optical correlators (AOC) while maintaining high-speed operation. However, neither the OEC nor the AOC in their conventional configurations can detect targets that have been rotated or scaled relative to a reference. This can be addressed by using a polar Mellin transform (PMT) pre-processing step to convert input images into signatures that contain most of the relevant information, albeit represented in a shift, scale, and rotation invariant (SSRI) manner. The PMT requires the use of optics to perform the Fourier transform and electronics for a log-polar remapping step. Recently, we demonstrated a pipelined architecture that can perform the PMT at a speed of 720 frames per second (fps), enabling the construction of an efficient opto-electronic PMT pre-processor. Here, we present an experimental demonstration of a complete HOC that implements this technique to achieve real-time and ultra-fast SSRI target recognition for space situational awareness. For this demonstration, we make use of a modified version of the HOC that makes use of Joint Transform Correlation , thus rendering the system simpler and more compact.

## 1. INTRODUCTION

As a fundamental part of space situational awareness, real-time target surveillance has attracted significant research interest. Traditionally, digital computational techniques have been implemented for target recognition thanks in part to their compatibility, widespread availability, and built-in re-programmability. However, due to the difficulty of dealing with two-dimensional data, the accuracy and speed of such systems is not optimal when compared to state of the art analog computation. Recent advancements in machine learning have shown progress in addressing these challenges, though they remain constrained by the low scalability of the digital domain. The algorithm YOLOv10, for example, has been developed for real-time object detection with a state-of-the-art convolutional neural network, achieving high detection accuracy with a latency time of 10.7 ms for recognizing colored images with a resolution of 384 x 640 pixels [1].

Optical signal processing systems have an inherent advantage in this regard, as the size of an image has little impact on the processing time due to the use of analog mechanisms to perform the required computations [2–4]. Optical correlators are of particular interest for space situational awareness. These systems use converging lenses to produce the Fourier transform (FT) of two input images, which can then be multiplied together and FT'd to produce a two-dimensional spatial cross-correlation [5]. Because the correlation is achieved through the mere propagation of light, the processing time is largely unaffected by the content or complexity of the images, operating at the speed it takes for light to propagate through them. The quality of the result is also largely independent of the complexity of the images as long as the laser, lenses and sensors are of high quality. However, these all-optical correlators are typically constrained by the speed at which the inputs can be updated. Additionally, they typically employ holographic filters or nonlinear materials for the multiplication step, which make real-world implementations difficult and fragile. We have previously shown how hybrid opto-electronic correlators (HOCs) may overcome some of these issues by replacing fragile materials with focal plane arrays (FPAs) and electronic processing [6]. While these types of correlators are still limited by the use of slow spatial light modulators (SLMs), we recently demonstrated a technique to improve the operating speed by incorporating a holographic image database, eliminating the need for ultrafast SLMs [7] for accessing the reference data base. This technique shows great advantages in recognizing objects from a large existing database, with images being stored in and retrieved from holographic memory discs.

Two-dimensional spatial optical correlations are inherently shift invariant due to the properties of the FT, yet they are incapable of detecting targets that have undergone scaling or rotation relative to the reference if no additional processing is performed. This impediment can be overcome by first pre-processing the input images to convert them into signatures that contain most of the relevant information, albeit represented in a shift, scale, and rotation invariant (SSRI) manner. We recently demonstrated that the polar Mellin transform (PMT) is an excellent candidate for such a pre-processing, as it can be computed as the log-polar transform (LPT) of the magnitude of the FT, and so can also take advantage of Fourier optics to enhance its speed. It is thus possible to construct an opto-electronic PMT pre-processing (OPP) stage that operates at the same speed as the rest of the correlation system [8,9].

Here, we propose a hybrid opto-electronic joint transform correlator (HOJTC), which is a simpler and more compact version of the conventional hybrid opto-electronic correlator (HOC), employing high-speed SLMs and integrated with the OPP stage, to realize an SRRI image correlator for real time object recognition. This system can perform target recognition at a speed of up to 720 fps, which corresponds to a response time of 1.4 milliseconds.

The rest of the paper is organized as follows. Section 2 provides an overview of the PMT and the scheme of the SSRI HOJTC. Detailed operation flowcharts are discussed in section 3. Section 4 presents experimental results that demonstrate a real-time HOJTC. Section 5 concludes this paper and gives our prospective insights.

## 2. SSRI REAL-TIME TARGET RECOGNITION VIA HOJTC

The OPP stage is a crucial image pre-processing part of this SSRI real-time target recognition system. We recently demonstrated that a pipelined architecture can produce the LPT at a speed of around one millisecond per image [9]. A diagram of the OPP system is shown in Fig.1. It works as follows. First, the original image is projected on a high-speed SLM which is placed at the focal plane of a lens such that its 2D spatial FT is produced at the opposite focal plane. A high-speed FPA connected to a computer via a PCI-express interface detects the resulting intensity of the FT. The computer captures the FT frame and uses a parallelized script to reorder the signal pixels according to the protocol for the LPT. Since the LPT depends exclusively on the coordinate system and not on the pixel values themselves, the input-output coordinate relationship of the LPT is pre-calculated and stored in memory to avoid real-time computations. Finally, the resulting signal is projected on a high-speed output SLM which itself functions as the input SLM for an HOJTC. The system maintains synchronicity through a series of hardware and software triggers. We describe the details of these steps below.

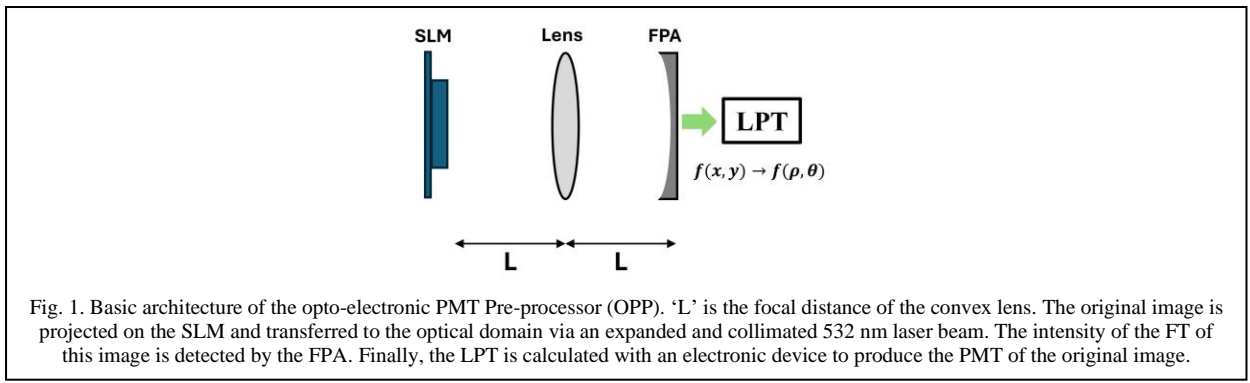

Fig. 1. Basic architecture of the opto-electronic PMT Pre-processor (OPP). 'L' is the focal distance of the convex lens. The original image is projected on the SLM and transferred to the optical domain via an expanded and collimated 532 nm laser beam. The intensity of the FT of this image is detected by the FPA. Finally, the LPT is calculated with an electronic device to produce the PMT of the original image.

The PMT is the LPT of the magnitude of the FT'd image, retaining most of the relevant information as the original image in a way that enables the correlation with shift, scale, and rotation differences [10]. After choosing coordinates such that the center of the FPA is the zero point ($x=0$, $y=0$), the calculation of the LPT starts as follows. First, the image is mapped to a different set of Cartesian coordinates where the horizontal axis is denoted as $r$ and the vertical axis is defined as $\theta$. The polar transform maps the ($x$, $y$) coordinates from the original image to ($r$, $\theta$) coordinates in the new image according to the relation $r = \sqrt{x^2 + y^2}$, and $\theta = \arctan(y/x)$. As such, the value of r ranges from 0 to a maximum positive value, while $\theta$ ranges from 0 to $2\pi$. Subsequently, the $r$ coordinate in the new image is rescaled to new coordinate denoted as $\rho$, based on the following expression: $\rho = \ln(r/r_0)$, where $r_0$ is a small distance, to be chosen judiciously. The mapping process is restricted to the domain of $r \geq r_0$, thus ensuring that the minimum value of $\rho$ is zero. In effect, this restriction of the domain represents blocking the DC component at the center of the FT'd image. This blocking is an essential aspect of the PMT process. In practice, this forced elimination of the DC component serves to enhance the prominence of the image features. The choice of the value of $r_0$ determines the extent of the low spatial frequency (2D) components that can be safely discarded without affecting the correlation process significantly.

Because the shift information for the image is encoded within the phase of its FT, it follows that upon the detection of the intensity of the FT, the shift is eliminated, providing shift invariance naturally. In scenarios where the object is rotated, the corresponding FT is also rotated by the same amount, φ, resulting in a linear shift φ along the $\theta$-axis in the PMT. Given that $\theta$ ranges from 0 to $2\pi$, the vertical axis of the PMT for a rotated image can be divided into two neighboring segments, one exhibiting a vertical width of φ and the other with a vertical width of $2\pi$-φ. Hence, rather than generating a single correlation peak, a rotated image would yield two correlation peaks, corresponding to the two segments. The separation of these peaks would be directly proportional to φ. In instances where the object is scaled by a factor of $\alpha$, the corresponding FT will be scaled by a factor of $1/\alpha$. Since $\rho(r/r_0\alpha) = \ln(r/r_0) - \ln(\alpha)$, this would result in a linear shift along the $\rho$-axis by an amount $\ln(\alpha)$. Thus, the value of α can be inferred from this shift. In some situation, one may be interested in recovering the lost shift information. In that case, the original image can be programmatically modified to eliminate the shift and scaling as per the measured valued. A subsequent correlation can be performed directly, i.e., without carrying out the PMT processing. The amount of shift for the image can then be inferred from the shift in the output correlation peak.

The all-optical JTC is discussed in detail in reference [5]. The HOJTC combines features of the HOC [6–9] with those of the JTC. The resulting architecture of the HOJTC, augmented by the OPP system, is illustrated schematically in Fig.2. First, a single SLM simultaneously projects the query and reference images after they have been pre-processed by the OPP, directing them towards a biconvex lens that produces the two overlapping FTs at its output plane. An FPA then detects the intensity of the interference between the FTs of the two input images. The resulting digital electronic signal contains two intensity terms, and two terms with the products of each FT with the conjugate of the other. This digital signal is then projected on an output SLM and FT'd using another biconvex lens. An FPA at the output plane detects a signal that contains the two-dimensional cross-correlation of the input images. The result will also contain additional signal from the intensity terms in the original FPA signal. We recently showed a technique by which this can be eliminated, forming a balanced JTC [11].

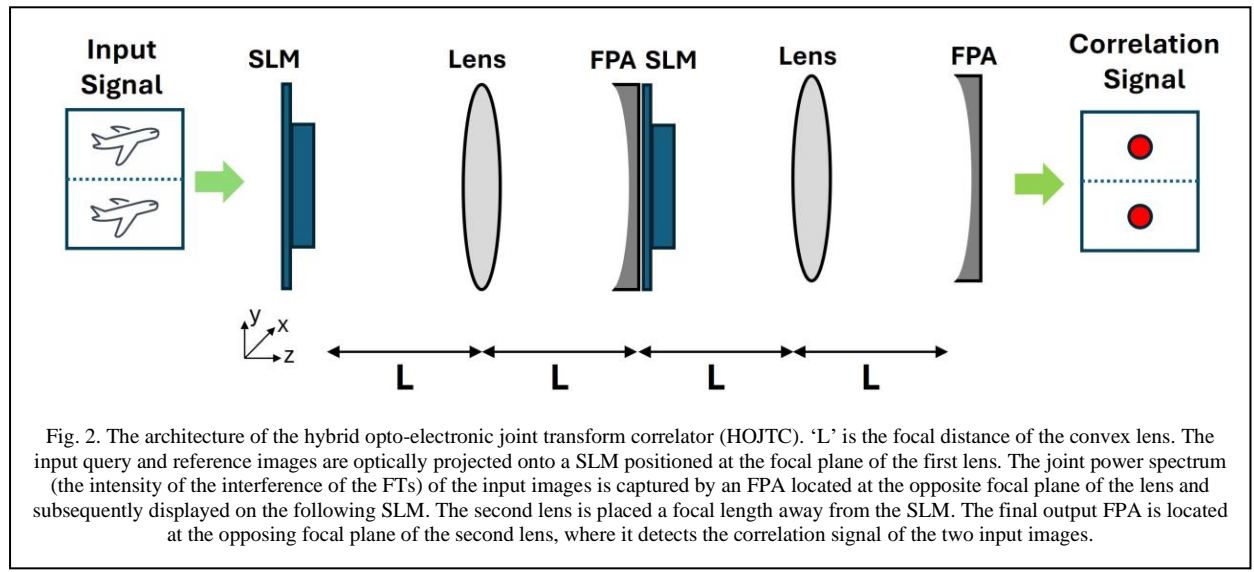

Fig. 2. The architecture of the hybrid opto-electronic joint transform correlator (HOJTC). 'L' is the focal distance of the convex lens. The input query and reference images are optically projected onto a SLM positioned at the focal plane of the first lens. The joint power spectrum (the intensity of the interference of the FTs) of the input images is captured by an FPA located at the opposite focal plane of the lens and subsequently displayed on the following SLM. The second lens is placed a focal length away from the SLM. The final output FPA is located at the opposing focal plane of the second lens, where it detects the correlation signal of the two input images.

## 3. SSRI HYBRID OPTO-ELECTRONIC JOINT TRANSFORM CORRELATOR

In the previous section, we discussed the scheme of the SSRI HOJTC. In summary, it combines an OPP stage with the HOJTC. As evident from Fig.1 and Fig.2, this correlation system employs three identical opto-electronic units, each of which can rapidly produce and detect the FT of the input images. Such a unit is composed of one SLM, one lens and one FPA. Here we propose two different system designs for the SSRI HOJTC. In one design (the full speed design), three such units work independently, yielding the maximum speed of operation. In the other (the single unit design), only one unit is employed for the entire correlation process, thus making it very compact, albeit at the expense of a reduction in the overall speed of operation. These are described in Section 3.1.

To realize optimal operational efficiency for the SSRI HOJTC, the device components, the data transmission techniques, and the system synchronization are of great importance. The device components are discussed in detail in reference [9], where we report an OPP stage working at a speed of 2ms per frame. Here we briefly describe the devices and report an experimentally demonstrated operation speed of 720 fps within a fully functional correlator system. The FPA utilized is a XIMEA Luxima LUX19HS monochromatic CMOS camera capable of functioning at over 2200 fps with full HD (1920x1080) resolution. This device interfaces with a computer via the PCI-express Gen3 standard, achieving a throughput of 64 Gbit/s. Higher framerates are possible when selecting a smaller Region of Interest (ROI), yet the lack of a high-speed commercial SLM remains a bottleneck for this architecture. Within our design framework, a Texas Instruments DLP471te digital micromirror device (DMD) SLM is selected, which operates at a rate of 240 fps for full-color (RGB888) frames with full HD resolution. It interfaces with a computer via the HDMI 2.1 standard. DMD SLMs operate by physically tilting their micromirrors about an axis, whereby for each pixel either all of the input light is directed towards the output, or none of it is. This binary operation is extended to monochromatic grayscale by pulse-width modulation, and further extended to RGB by sequentially displaying three monochrome frames that encode the red, green, and blue information, respectively. Because of this sequential operation, it can effectively be treated as a monochromic display with a speed of 720 fps.

Synchronization between the SLM and the FPA is essential; while each individual frame is being displayed, the FPA must start the exposure, capture an image, and transmit the result to a computer. Because the DMD SLM is designed for sequential RGB operation, it produces an electronic enable signal for each red, green, and blue segment of the frame. This is intended to control colored LEDs which are designed with specific durations to ensure perceptible color accuracy to the human eye. A Microchip ATSAMD21A microcontroller is used to detect the rising edge of either of the LED enable signals and subsequently send a trigger signal with a duration of ~300 μs to the camera following a predetermined time delay of ~200 μs. The general-purpose input/output (GPIO) of the camera is configured to recognize the rising edge of this signal to trigger the start of image capture. A lens is placed between the SLM and the camera, enabling the camera to sequentially capture the FT of the input images. The Ximea camera can also produce a capture completion signal, which can be used by the computer to start projecting on the subsequent SLM. This cascaded architecture has not yet been employed, and instead a single-module version is reported on here. This is further discussed in section 3.1 below. It should also be noted that the SLM expects an RGB frame through its HDMI connection, and so three independent grayscale frames must be encoded into a single RGB frame before transmission.

The timing of the devices is shown in Fig.3. The selected SLM operates at 240 Hz, full color, and full HD, which allows the display of 720 frames each second with a resolution of 1920x1080 pixels. This is equivalent to ~1.4ms allocated for each individual frame, during which time it is necessary to capture its FT, process, and display it. Given that this SLM was calibrated for uniform color perception by humans, the duty cycle for each red, green, and blue frame are 35.48%, 44.29% and 18.73%, respectively, which are labelled as $t_{red}$, $t_{green}$ and $t_{blue}$.

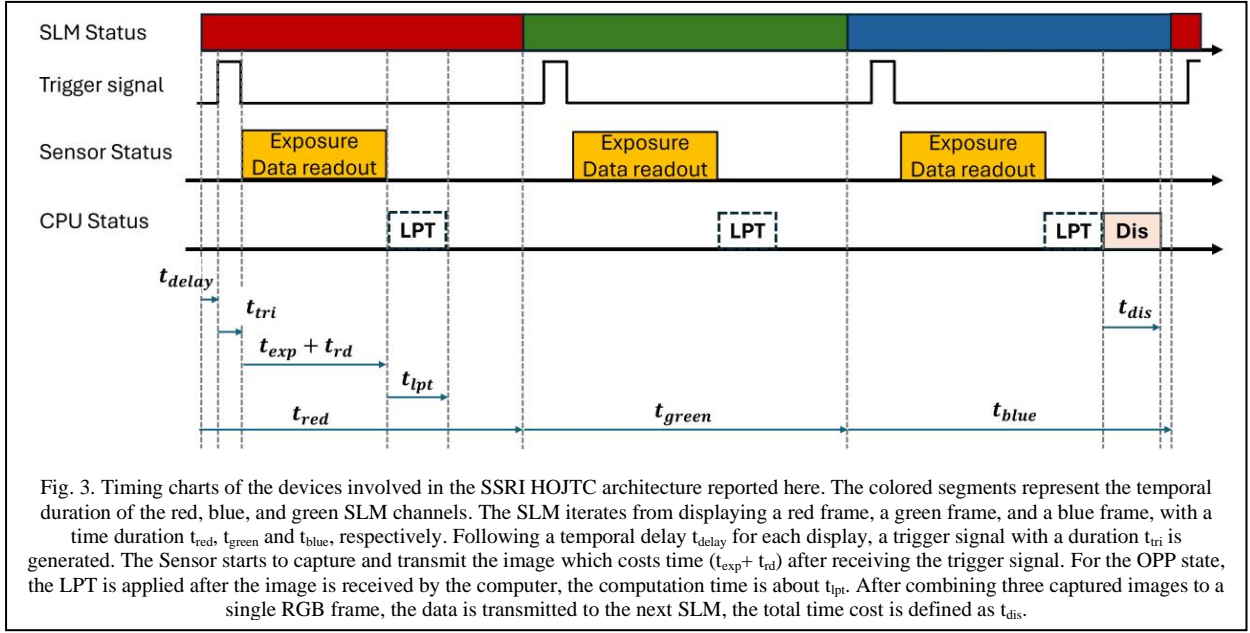

Fig. 3. Timing charts of the devices involved in the SSRI HOJTC architecture reported here. The colored segments represent the temporal duration of the red, blue, and green SLM channels. The SLM iterates from displaying a red frame, a green frame, and a blue frame, with a time duration $t_{red}$, $t_{green}$ and $t_{blue}$, respectively. Following a temporal delay $t_{delay}$ for each display, a trigger signal with a duration $t_{tri}$ is generated. The Sensor starts to capture and transmit the image which costs time ($t_{exp}+ t_{rd}$) after receiving the trigger signal. For the OPP state, the LPT is applied after the image is received by the computer, the computation time is about $t_{lpt}$. After combining three captured images to a single RGB frame, the data is transmitted to the next SLM, the total time cost is defined as $t_{dis}$.

Once a frame is being displayed, the microcontroller will wait a time delay of $t_{delay}$ and emit a trigger signal to the camera. The pulse duration is defined as $t_{tri}$, which must exceed a threshold of 300 µs to ensure the camera can accurately detect a valid rising edge. The sensor will start the exposure and transmit the image from the sensor RAM to the computer RAM when it detects the rising edge of the trigger signal. Following this operation, the sensor will revert to a *ready* state in anticipation of the next trigger. The exposure duration is set to be a constant $t_{exp}$, which can be adjusted according to the laser's power. This time is different for the OPP HOJTC stages, with the HOJTC typically necessitating a much longer exposure. This requirement arises from the observation that a PMT'd image exhibits a lower overall brightness than the original image thanks to the DC block, thus requiring a lengthened exposure time to achieve full exposure of the FT results of the input, assuming the laser power is equal in both stages. The duration allocated for data readout is denoted as $t_{rd}$, which is contingent upon the size of the ROI. In our experimental setup, the size of the FT was small enough to allow for the FPA ROI to be configured to a resolution 320x 320 pixels. Once the data is transferred to a computer, the LPT is performed in parallel with a multi-core CPU. After three independent frames are captured and processed, they are merged into one RGB frame for display on the subsequent SLM, with the time for this operation defined as $t_{dis}$. For the results reported here, the frame rendering was controlled using OpenGL. As soon as the RGB frame is ready, it is rendered within a spatial window with the same size as the frame created by OpenGL. More precise timing information is discussed in section 4.

### 3.1 FULL SPEED AND SINGLE UNIT SSRI HOJTC DESIGNS

As previously mentioned, the SLM is restricted to receiving RGB-encoded frames; therefore, it is necessary to process sets of three distinct monochromatic frames in order to subsequently combine them into a singular RGB frame that is then transmitted via HDMI. We propose two process-designs to optimize the speed and performance of this hybrid opto-electronic correlator system. The first design (denoted as the **full speed design**) flowchart is shown in Fig.4. Here, three different query images are allocated to different colored frames and concurrently displayed on the SLM. In this scheme, each physical stage of the SSRI HOJTC is dedicated to different steps in the process; the first stage is the OPP, the second stage is the correlation input, and the third stage is the correlation output. Each of the three input images is assigned to its own color channel, which is maintained throughout the stages. That is to say, the first image is always assigned to the red channel, the second to the green, and the third to the blue.

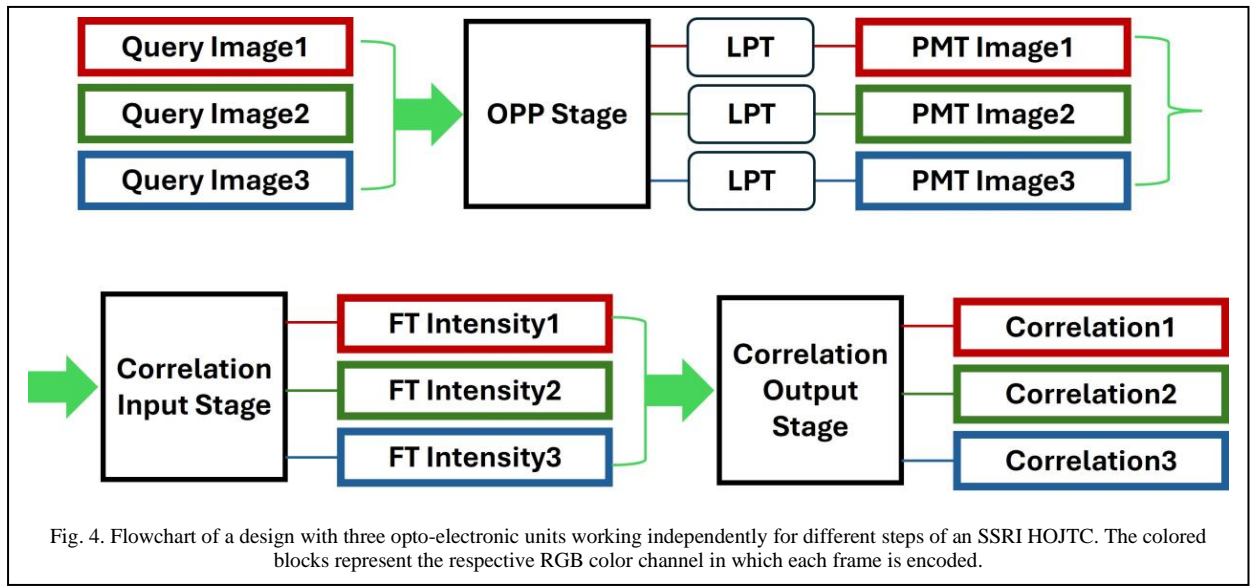

Fig. 4. Flowchart of a design with three opto-electronic units working independently for different steps of an SSRI HOJTC. The colored blocks represent the respective RGB color channel in which each frame is encoded.

As discussed before, different color frames have different display durations. In addition to this, they also possess different levels of color illumination, which correspond to the maximum grayscale amplitude of the frame. For the SLM used here, the default illumination for red, green, and blue are set to be 67, 259, and 23, respectively, with the ratio between the maximum and minimum illumination intensities being approximately 11:1. This disparity complicates the normalization of the frame amplitude, thereby posing challenges to achieving uniform performance across each frame, as is essential for the construction of a full-speed 720 fps correlator. For simplicity, the experiments reported here are implemented utilizing a singular opto-electronic FT unit that first acts as the OPP, then as the correlation input stage, and finally as the correlation output stage.

As was previously mentioned, the exposure duration for the OPP and correlation segments can be significantly different. This presents an opportunity for an alternative architecture (the **single unit design**) that only requires one opto-electronic FT unit, whereby the default illumination ratios for each color channel can be used to our advantage. To this end, each of the three opto-electronic FT stages can be assigned to one of the three color frames. The OPP stage typically requires a shorter exposure time, and so it may be assigned to the blue frame, which has the minimum illumination. The correlation input stage can be assigned to the red frame, which has an intermediate projection time, and the output correlation stage can be assigned to the green frame, which has the longest projection time. In this architecture, the green frame will always contain the correlation output, but the overall SSRI correlation speed is limited to 240 fps. The flowchart of the single unit design is presented in Fig.5. In the initial stage, only the first query image is exhibited on the blue frame, and the LPT is processed immediately after the opto-electronic FT unit to produce the PMT image. Then the received PMT image is assigned to the red frame, while the blue frame is concurrently displaying the second query image. As the PMT image traverses the FT unit, it generates the intensity of the FT of the PMT query and reference image, which will be assigned to the green frame when it is displayed on the following SLM. In the third processing phase, the red frame will present the PMT of the second query image, the green frame will carry FT intensity of the PMT of the first image, and the blue frame will represent a third query image. The FPA will detect three FT'd images in sequence. The first frame is the FT intensity of the PMT of the second query and reference image, the second frame presents the correlation signal of the first query image with the reference image. The third frame depicts the PMT of the third image.

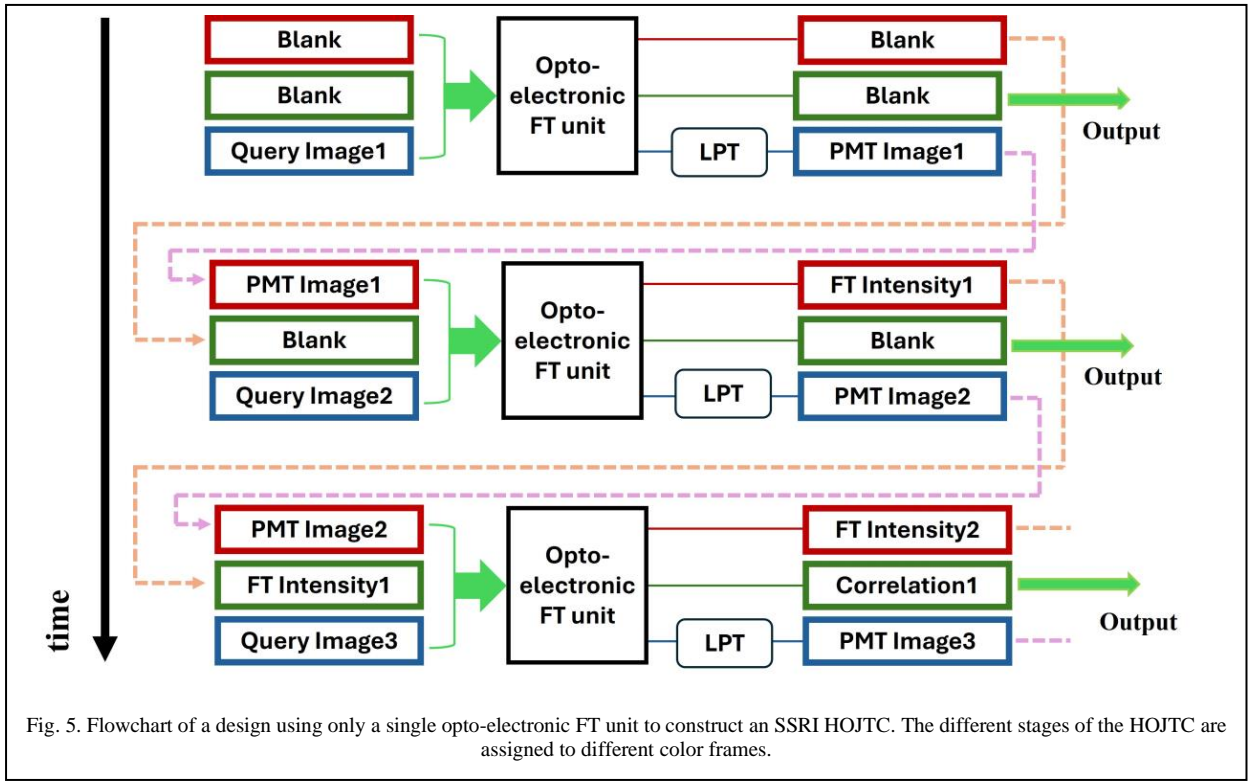

Fig. 5. Flowchart of a design using only a single opto-electronic FT unit to construct an SSRI HOJTC. The different stages of the HOJTC are assigned to different color frames.

## 4. EXPERIMENTAL RESULTS

The SSRI HOJTC designs described in the precious section were constructed and tested using a single opto-electronic FT unit for simplicity. We evaluated the speed and performance of both designs. The experimental setups are shown in Fig.6. Fig.6(A) shows a single opto-electronic FT unit. Centrally located is a convex lens with a focal length of 100 millimeter. The SLM is placed at the focal plane of this lens, while the FPA is positioned on the opposite focal plane. In Fig.6(B), the microprocessor used for generating a trigger signal is shown. This unit receives three LED enable signals from the SLM, which are subsequently converted into an appropriately timed singular trigger output for the FPA.

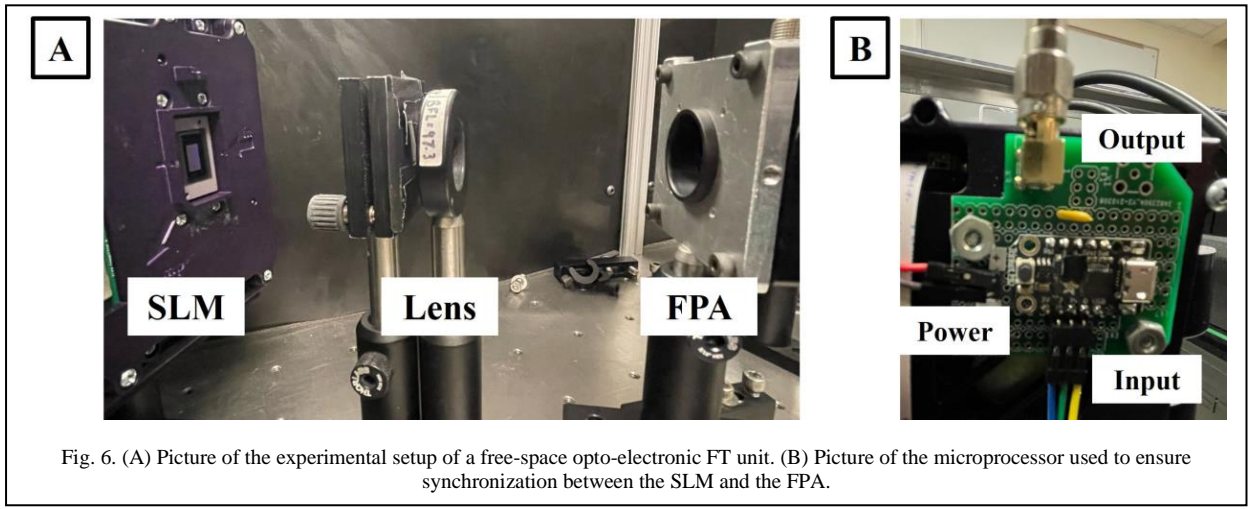

Fig. 6. (A) Picture of the experimental setup of a free-space opto-electronic FT unit. (B) Picture of the microprocessor used to ensure synchronization between the SLM and the FPA.

In what follows, we report two different sets, to be denoted as **Set A** and **Set B**, of results, both using the same physical apparatus shown in Fig.6. However, the coding of the information was done differently for each set of results. For Set A, we used a coding approach that would be used if three different opto-electronic FT units were used (corresponding to the **full speed design**) for maximizing the overall operating speed of the SSRI HOJTC. For Set B, we used a coding approach that would be used if only one opto-electronic FT unit were used (corresponding to the **single unit design**) for a compact implementation of the SSRI HOJTC, at the expense of a reduction in the operating speed.

The experimental results obtained under Set A are presented in Fig.7. As noted above, the full speed design requires three sets of the opto-electronic FT units; however, the performance of each unit can be independently evaluated to yield an overall metric. As such, the speed and performance of each section was independently tested. For these tests, the reference image consisted of a square with a rounded four-pointed star inside. The first query image is identical to the reference image and corresponds to an autocorrelation, the second query image is rotated 45 degrees with respect to the reference, and the third query image is scaled by a factor of 0.9. These images are normalized to have the same total optical power. The resulting composite RGB frame is shown in Fig.7(A0) and serves as the input to the first SLM. The corresponding PMT images for each input query image are presented in Fig.7(A1), (A2), and (A3). The exposure time was set to 10 µs per monochrome frame. The FT image resolution is 320 x 320, with the cutoff $r_0$ defined as 16 pixels. In these results, the effect of scaling and rotation are clearly visible, where a change in scale results in a linear horizontal shift, and a change in angle results in a vertical shift (modulo 2π). This is also especially visible in the combined RGB frame that was generated as a combination of the three monochrome LPT frames, which is shown in Fig.7(B0). The experiments were conducted for 2000 iterations, averaging the time that takes for a single round of the OPP process, which is defined as the time between the start of the FPA capture for the first monochrome frame and the SLM display of the processed RGB-encoded PMT image, divided by three. The experiments recorded an average opto-electronic PMT speed of 719.62 fps, with each monochrome digital LPT taking 18 microseconds. The display time $t_{dis}$ is measured at 794 µs for the combined RGB-encoded PMT frame. This can match the full operational speed of the SLM, which is ~1.389 ms per frame.

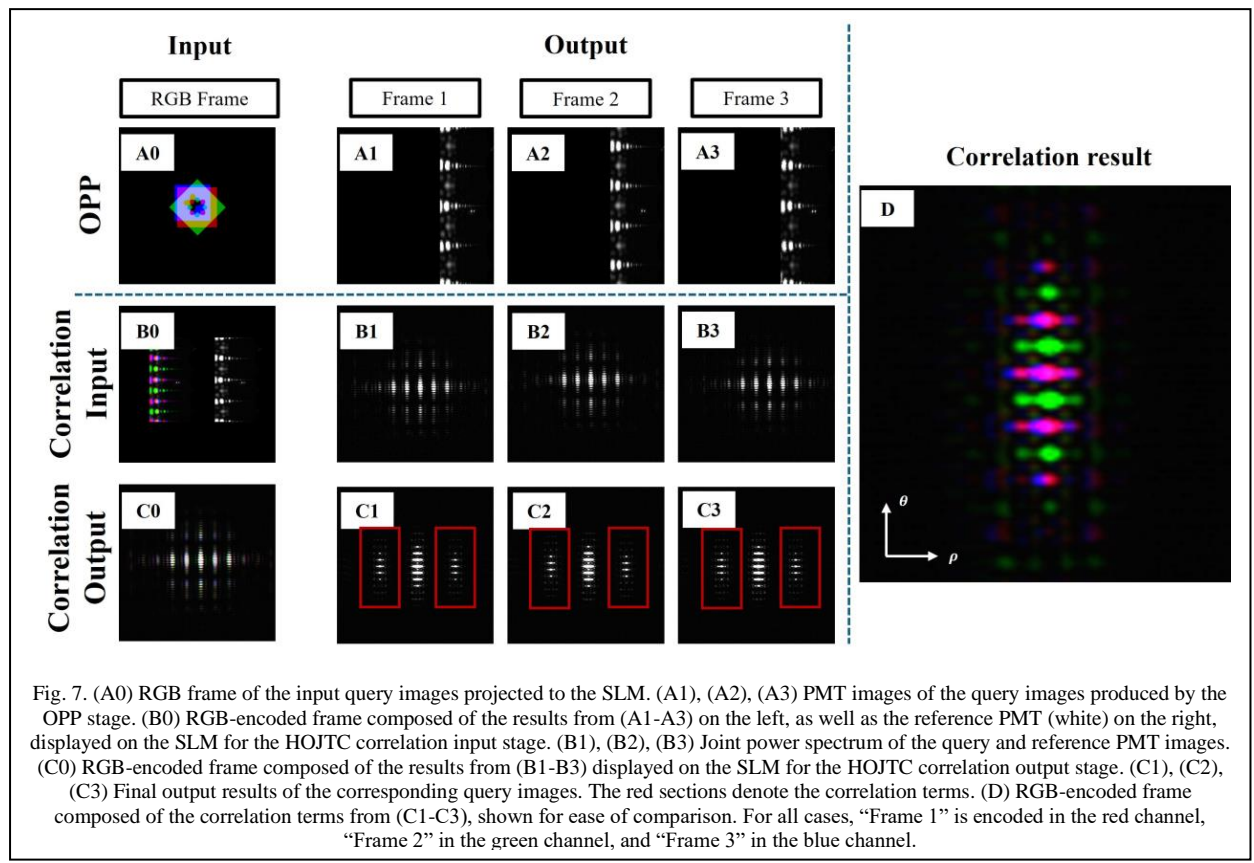

Fig. 7. (A0) RGB frame of the input query images projected to the SLM. (A1), (A2), (A3) PMT images of the query images produced by the OPP stage. (B0) RGB-encoded frame composed of the results from (A1-A3) on the left, as well as the reference PMT (white) on the right, displayed on the SLM for the HOJTC correlation input stage. (B1), (B2), (B3) Joint power spectrum of the query and reference PMT images. (C0) RGB-encoded frame composed of the results from (B1-B3) displayed on the SLM for the HOJTC correlation output stage. (C1), (C2), (C3) Final output results of the corresponding query images. The red sections denote the correlation terms. (D) RGB-encoded frame composed of the correlation terms from (C1-C3), shown for ease of comparison. For all cases, "Frame 1" is encoded in the red channel, "Frame 2" in the green channel, and "Frame 3" in the blue channel.

The RGB-encoded PMT image is projected onto the SLM as shown in Fig.7(B0), the left portion of the frame contains the query images, while the right contains the reference images, which appear white as the three frames use the same reference. If different reference images were desired, they would also be encoded in RGB. This serves as the input for the HOJTC correlation stage. The exposure time of the FPAs in this stage is set to be 200 µs. The total time of exposure and data readout is ~0.21ms per monochrome frame. The subsequent FPA detects the joint power spectrum of the inputs, as shown in Fig.7(B1), (B2), and (B3). This process is also tested for 2000 rounds, yielding an average speed of ~719.58 fps. The results joint power spectra of the three frames are again combined into an RGB-encoded frame to serve as the input to the final stage, as shown in Fig.7(C0), which produces the output correlation results. The outputs from the FPA are shown in Fig.7 (C1), (C2), and (C3), where the correlation terms are outlined in red. Fig.7 (D) shows the correlation results overlaid in an RGB frame to highlight the differences between the three correlations. The red result contains an autocorrelation, and so serves as a reference. The green result has a 45° rotation, and so appears vertically shifted (modulo $2\pi$) with respect to the red result. The blue result is scaled by a factor of 0.9, and so appears horizontally shifted to the left with respect to the red result. These results are exactly as expected for an SSRI hybrid opto-electronic correlator, and show that the RGB-encoding has no impact on the quality of the data. A total of 2000 iterations are conducted for this correlation process, maintaining an average correlation speed of ~719.55 fps, indicating that the three processes synchronize in speed and can be integrated to achieve an overall speed over 719 fps.

The experimental results obtained under **Set B** are shown in Fig.8, where the featured RGB input frame and corresponding output results are displayed. The process is tested for 2000 rounds to measure an average processing time. The exposure time of the FPA in this design is set to be 200 µs, where the default color-correction of the SLM is used to obtain a more uniform exposure, as previously explained. Fig.8(A0) shows the RGB-encoded input frame, which contains the original input image in the blue channel, the OPP-output PMT/ HOJTC correlation input in the red channel, and the HOJTC joint power spectrum in the green channel. These three frames are shown separately in grayscale in Fig.8 (A1), (A2), and (A3). The FPA detects the resultant outputs, shown in Fig.8 (B1), (B2), and (B3), and feeds them back into the system as needed, where the green frame's output contains the correlation data. The processing time for each monochrome frame was measured to be ~719.67 fps on average. However, because three such iterations are required to obtain the correlation signal for a single image, the real-world operational speed of this compact correlator is ~239.89 fps on average.

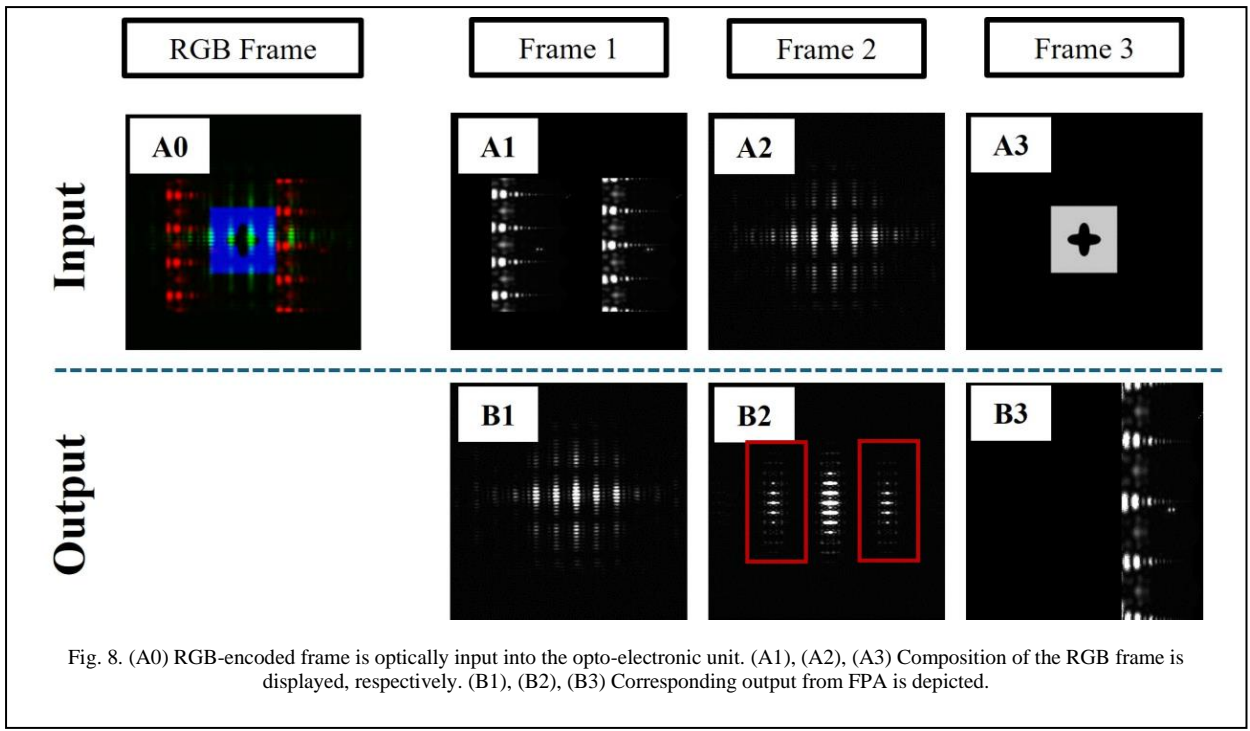

Fig. 8. (A0) RGB-encoded frame is optically input into the opto-electronic unit. (A1), (A2), (A3) Composition of the RGB frame is displayed, respectively. (B1), (B2), (B3) Corresponding output from FPA is depicted.

## 5. CONCLUSIONS AND OUTLOOK

A high-speed SSRI HOJTC was demonstrated and constructed, with off-the shelf components, enabling a viable solution to real-time target recognition that may be implemented today for space situational awareness. The 720fps operating speed for images with a resolution of 1920x 1080 is, to the best of our knowledge, the fastest experimentally demonstrated SSRI image correlation. For simplicity, we demonstrate a compact design with only a SLM, a lens and an FPA which form an opto-electronic unit designed to build up a target recognition correlator that runs at a speed of 240 fps. We have illustrated that by integrating three such units and achieving precise synchronization among them, it is feasible to reach a speed at 720 fps, which is almost a magnitude faster than the leading computation method. The continuous improvements of high-speed SLMs and FPAs will only improve the functionality of these devices, offering an enticing future in robust hybrid opto-electronic high-speed target recognition systems for space situational awareness. As an ultra-fast computational unit for image correlation calculation, this design can further be applied as a computational accelerator for convolution neural networks at target recognition tasks.

## 6. FUNDING

The work reported here was supported by AFOSR grant No. FA9550-18-01-0359.